\documentclass[12pt]{article}
\usepackage{a4wide,amsmath,amsthm,epsfig}
\usepackage{amsmath,amsthm}
\usepackage{hhline}
\usepackage{amsfonts}

\newcommand{\url}[1]{}

\newcommand{\Qx}{ \mathbb{Q} }
\newcommand{\Ex}{ \mathbb{E} }

\newcommand{\vertvect}[2]{\left[ \begin{array}{c} #1 \\ #2 \end{array}
\right]}

\newcommand{\cds}{\mbox{CDS}}
\newcommand{\prcds}{\mbox{\small PRCDS}}

\newcommand{\barP}{\bar{P}}
\newcommand{\pcds}{\Pi\mbox{\tiny RCDS}}
\newcommand{\pprcds}{\Pi\mbox{\tiny PRCDS}}

\newcommand{\lgd}{\mbox{L{\tiny GD}}}
\newcommand{\rec}{\mbox{R{\tiny EC}}}
\newcommand{\cmcds}{\mbox{CDS\tiny{CM}}}
\newcommand{\Rtil}{\widetilde{R}}

\title{
{\normalsize Shortened version published in Risk Magazine, June 2006 issue.}\\
{\small Related paper in ``Credit Risk: Models, Derivatives and Management", Taylor \& Francis
}\\ --- \\
{\large \bf Constant Maturity Credit Default Swap Pricing\\
with Market Models
\thanks{The author is grateful to
Aur\'elien Alfonsi for fundamental comments on an earlier draft
and to Massimo Masetti and  Massimo Morini for helpful
suggestions. Massimo Masetti contributed also with  numerical
examples and tests. The interest in pricing this product
originated from communication with Marco Salcoacci in Munich,
whose timing in grasping market sentiment has played a fundamental
role }}
}
\author{Damiano Brigo
   \\ Credit Models,
Banca IMI \\
Corso Matteotti 6, 20121 Milano, Italy \\
{\small  {\tt  http://www.damianobrigo.it}} 
}
\date{{\small First Version: October 20, 2004. \\ Published in the ``Social Science Research Network" on December 21, 2004.}}
\newtheorem{theorem}{Theorem}[section]
\newtheorem{proposition}[theorem]{Proposition}

\newtheorem{definition}[theorem]{Definition}

\newtheorem{remark}[theorem]{Remark}

\begin{document}
\maketitle

\begin{abstract}
In this work we derive an approximated no-arbitrage market
valuation formula for Constant Maturity Credit Default Swaps
(CMCDS). We move from the CDS options market model in Brigo
(2004), and derive a formula for CMCDS that is the analogous of
the formula for constant maturity swaps in the default free swap
market under the LIBOR market model. A ``convexity
adjustment"-like correction is present in the related formula.
Without such correction, or with zero correlations, the formula
returns an obvious deterministic-credit-spread expression for the
CMCDS price. To obtain the result we  derive a joint dynamics of
forward CDS rates under a single pricing measure,  as in Brigo
(2004). Numerical examples of the ``convexity adjustment" impact
complete the paper.
\end{abstract}

{\bf Keywords:} CDS Options, CDS Options Market Model, Constant
Maturity CDS, Convexity Adjustment, Participation Rate, CDS rates
volatility, CDS rates correlation.

\newpage

\tableofcontents

\newpage

\section{Introduction}
Constant Maturity Credit Default Swaps (CMCDS) are receiving
increasing attention in the financial community. In this work, we
aim at deriving an approximated no-arbitrage market valuation
formula for CMCDS. We move from the CDS options market model in
Brigo (2004), and derive a formula for CMCDS that is the analogous
of the formula for constant maturity swaps in the default free
swap market under the LIBOR market model.

In Brigo (2004) the focus is somehow different, since in that
paper we derive CDS option prices for CDS payoffs given in the
market and for some postponed approximated CDS payoffs. We do so
by means of a rigorous change of numeraire technique following
Jamshidian (2002) and based on a result of Jeanblanc and Rutkowski
(2000). We also establish a link with callable defaultable
floaters.

This paper starts by recalling again one alternative expression
for CDS payoffs, stemming from different conventions on the
premium flows and on the protection leg.  We briefly introduce CDS
forward rates, postponing the detailed definitions. We define
CMCDS and give immediately the main result of the paper, the
approximated pricing formula, in terms of CDS forward rates and of
their volatilities and correlations. We point out some analogies
with constant maturity swaps, showing that a ``convexity
adjustment"-like correction is present. Without such correction,
or with zero correlations, the formula returns an obvious
deterministic-credit-spread expression for the CMCDS price.

Once the main result has been described, we move to introduce the
formal apparatus that allows to prove it. We introduce rigorously
CDS forward rates. This leads to an investigation on the
possibility to express such rates in terms of some basic
one-period rates and to a discussion on a possible analogy with
the LIBOR and swap default-free models. We discuss the change of
numeraire approach to deriving a joint dynamics of forward CDS
rates under a single pricing measure, derivation that is only
hinted at in Brigo (2004) for deriving the Black-like formula for
CDS options. Through a drift-freezing approximation we then prove
the formula for CMCDS pricing and give some numerical examples
highlighting the role of the participation rate and of the
convexity adjustment.

\begin{remark}{\bf (How to use this paper)} The CDS Options market model is the
same as in Brigo (2004) and is based on a complex change of
numeraire involving two families of forward CDS rates. This
apparatus is reported in Section~\ref{onetworates} {\underline but
there is no need to go through it if one does not plan to test the
formula against Monte Carlo simulation}. As far as the formula
derivation is concerned, the reader may skip
Section~\ref{onetworates} and go directly to the much simpler
approximated model in Section~\ref{oneratesonly}.
\end{remark}

\section{Credit Default Swaps (CDS) and Constant Maturity CDS}\label{CDSdiffform}


We recall briefly some basic definitions for CDS's.

\begin{definition}{\bf (Credit Default Swap)}.
A CDS is a contract where the ``protection buyer" ``A" pays rates
$R$ (``$R$" stands for Rates) at times $T_{a+1},\ldots,T_b$ (the
``premium leg") in exchange for a single protection payment $\lgd$
(loss given default, the protection leg). ``A" receives the
protection leg by the ``protection seller" ``B" at the default
time $\tau$ of a reference entity ``C", provided that $T_a < \tau
\le T_b$. The rates $R$ paid by ``A" stop in case of default. We
thus have

\[
\hspace{-1cm} \boxed{\begin{array}{ccccc}
\mbox{Protection Seller} & \rightarrow & \mbox{ protection } \lgd \mbox{ at default $\tau_C$ of ``C" if in $[T_a,T_b]$} & \rightarrow & \mbox{Protection Buyer} \\
%
\mbox{``B"} & \leftarrow  & \mbox{ rate } R \mbox{ at }
T_{a+1},\ldots,T_b \mbox{ or until default } \tau_C & \leftarrow &
\mbox{``A"}
\end{array}} \]

This is called a ``running CDS" (RCDS) discounted payoff.
\end{definition}

We explicitly point out that we are assuming the offered
protection amount $\lgd$ to be deterministic. Typically
$\lgd=1-\rec$, where the recovery rate $\rec$ is assumed to be
deterministic and the notional is set to one.

Sometimes a slightly different payoff is considered for RCDS
contracts. Instead of considering the exact default time $\tau$,
the protection payment $\lgd$ is postponed to the first time $T_i$
following default. If the grid is three or six months spaced, this
postponement consists in a few months at worst.

We term ``Postponed payments Running CDS" (PRCDS) the CDS payoff
under the postponed formulation.  The advantage of the postponed
protection payment is that no accrued-interest term is necessary,
and also that all payments occur at the canonical grid of the
$T_i$'s. The postponed payout is better for deriving market models
of CDS rates dynamics, as we shall see shortly. It is also
fundamental in establishing the ``defaultable floater analogy" as
we have seen in Brigo (2004).

We denote by $\cds_{a,b}(t, R, \lgd)$ the price at time $t$ of the
above standard running CDS flows to the protection seller ``B". We
add the prefix ``PR" to denote the analogous price for the
postponed payoff. The pricing formulas for these payoffs depend on
the assumptions on interest-rate dynamics and on the default time
$\tau$. In general the CDS forward rate $R_{a,b}(0)$ for
protection in $T_a,T_b$ at initial time $0$ is obtained (in the
postponed case) by solving the equation $\prcds_{a,b}(0,
R_{a,b}(0), \lgd)=0$. We will give details on this later: now we
only say that the market provides quotes for $R_{0,b}(0)$'s for
increasing maturities $T_b$ (notice that $T_0 = 0$, so that only
``spot" CDS rates are quoted in the market). The $R_{a,b}(0)$ rate
makes the CDS contract fair at the valuation time. A special role
in our work will be covered by one-period rates $R_i(0) =
R_{i-1,i}(0)$ (protection in $[T_{i-1},T_i])$. These rates may
seem artificial but they can be easily computed from quoted spot
rates. An important role is also assigned to the corporate
zero-coupon bond price $\barP(0,T)$, which is the price at time
$0$ of one unit of currency made available by name ``C" at
maturity $T$ if no default occurs, and with no recovery in case of
early default. The corresponding default-free zero coupon bond is
denoted by $P(0,T)$.

We are now ready to introduce CMCDS's.

In a CMCDS with first reset in $T_a$ and with final maturity
$T_b$, protection $\lgd$ on a reference credit ``C" against
default in $[T_a,T_b]$ is given from a protection seller ``B" to a
protection buyer ``A". However, in exchange for this protection, a
``constant maturity" CDS rate is paid.

We know by definition that the fair rate to be paid at $T_i$ for
protection against default in $[T_{i-1},T_i]$ would be $R_i$. This
leads us to the following

\begin{remark} {\bf (A ``floating-rate" CDS).}
A contract that protects in $T_a,T_b$ can be in principle
decomposed into a stream of contracts, each single contract
protecting in $[T_{j-1},T_j]$, for $j=a+1,\ldots,b$, say with
protection payment $\lgd$ postponed to $T_j$ if default occurs in
$[T_{j-1},T_j]$. In each single period, the rate $R_j(T_{j-1})$
paid at $T_j$ makes the exchange fair, so that in total a contract
offering protection $\lgd$ on a reference credit ``C" in
$[T_a,T_b]$ in exchange for payment of rates $R_{a+1}(T_{a}),
\ldots, R_j(T_{j-1}),\ldots, R_{b}(T_{b-1})$ at times
$T_{a+1},\ldots,T_j,\ldots,T_b$ is fair, i.e. has zero initial
present value. This product can be seen as a sort of floating rate
CDS.
\end{remark}

However, in CMCDS's the rate that is paid at each period for
protection is not the related one-period CDS rate, as would be
natural from the above remark, but a longer period CDS rate.
Consider indeed the following

\begin{definition}\label{cmcdsdef} {\bf (Constant Maturity CDS).}
Consider a contract protecting in $[T_a,T_b]$ against default of a
reference credit ``C". If default occurs in $[T_a,T_b]$, a
protection payment $\lgd$ is made from the protection seller ``B"
to the protection buyer ``A" at the first $T_j$ following the
default time. This is called ``protection leg". In exchange for
this protection ``A" pays to ``B" at each $T_j$ before default a
``$c+1$--long" (constant maturity) CDS rate $R_{j-1,j+c}(T_{j-1})$
(times a year fraction $\alpha_j = T_j - T_{j-1}$), with ``$c$" an
integer larger than zero. Notice that for $c=0$ we would obtain
the fair ``floating rate" CDS above, whose initial value would be
zero.
\end{definition}

Given that $c>0$ in our definition, the value of the contract will
be nonzero in general, so that we have to find this value at the
initial time $0$ if we are to price this kind of transaction. We
face this task by resorting to the market model derived below.

The value of the CMCDS to ``B" is the value of the premium leg
minus the value of the protection leg. The protection leg
valuation is trivial, since this is the same leg as in a standard
forward start $[T_a,T_b]$ CDS. As such, it is for example equal to
\[ R_{a,b}(0) \sum_{j=a+1}^b \alpha_j \barP(0,T_j) = \sum_{j=a+1}^b \alpha_j R_j(0) \barP(0,T_j) . \]
This value has to be subtracted to the premium leg. The
non-trivial part is indeed computing the premium leg value at
initial time 0. Notice that the final formula can be implemented
easily on a spreadsheet, requiring no numerical apparatus.

\begin{proposition}\label{propmainrescmcds} {\bf (Main Result: An approximated formula for CMCDS)}
Consider the Constant Maturity CDS defined in
Definition~\ref{cmcdsdef}. The present value at initial time $0$
of the CMCDS to the protection seller ``B" is

\begin{eqnarray}\label{cmcdsvolscorr}
\hspace{-1cm} \cmcds_{a,b,c}(0,\lgd) = \sum_{j=a+1}^b \alpha_j
\barP(0,T_j) \left\{ \sum_{i=j}^{j+c}  \frac{\alpha_i
\barP(0,T_i)}{\sum_{h=j}^{j+c} \alpha_h \barP(0,T_h)} \cdot
\right. \\ \nonumber \left.
\widetilde{R}_i(0)\exp\left[T_{j-1}\sigma_i\cdot\left(
\sum_{k=j+1}^{i}\rho_{j,k}\frac{\sigma_k\widetilde{R}_k(0)}{\widetilde{R}_k(0)+\lgd
/\alpha_k}\right)\right]  - R_j(0)\right\}
\end{eqnarray}

where $R_k(0)$  are the one-period CDS forward rates for
protection in $[T_{k-1},T_k]$. These CDS forward rates can be
computed from quoted spot CDS rates $R_{0,k}(0)$ and corporate
zero coupon bonds $\barP(0,T_k)$ via

\begin{eqnarray*}
R_k(0) = \frac{R_{0,k}(0) \sum_{h=1}^{k} \alpha_h \barP(0,T_h) -
R_{0,k-1}(0) \sum_{h=1}^{k-1} \alpha_h \barP(0,T_h)}{\alpha_k
\barP(0,T_k)}  \\
\end{eqnarray*}

while $\Rtil_k(0)$ are approximations of the $R_k(0)$ (equal in
case of independence of interest rates and credit spreads) in
terms of corporate $\barP$ and default free $P$ zero coupon bonds
given by

\[  R_k(0) \approx \Rtil_k(0) = \lgd \frac{\barP(0,T_{k-1})
 P(0,T_k)/P(0,T_{k-1})- \barP(0,T_k)}{ \alpha_k\bar{P}(0,T_k)}\]
and where: $\sigma_k$ is the volatility of $R_k(t)$, assumed
constant (we deal with the time-varying volatility in the proof
below);

$\rho_{i,j}$ is the instantaneous correlation between $R_i$ and
$R_j$;

One-period forward CDS rates volatilities $\sigma_k$ can in
principle be stripped from longer period CDS volatilities,
similarly to how forward LIBOR rates volatilities can be stripped
from swaptions volatilities in the LIBOR model. This stripping is
made possible from an approximated volatility formula based on
drift freezing (formula (6.58) for the LIBOR case in Brigo and
Mercurio (2001)). Cascade methods are also available for this (as
in Brigo and Morini (2004)), although for the time being the only
available CDS options all have short maturities and the lack of a
liquid market discourages this kind of approach. For the time
being the above formula can be employed with stylized values of
volatilities to have an idea of the impact of the ``convexity
adjustments". Finally, one may consider using historical
volatilities and correlations in the formula as first guesses.

As a further remark we notice that, if not for the exponential
term (which vanishes for example when $\rho$'s are set to zero)
this expression would be, not surprisingly,

\begin{eqnarray}\label{cmcdsnotsur}
\hspace{-1cm} \cmcds_{a,b,c}(0,\lgd;\rho=0) = \sum_{j=a+1}^b
\alpha_j \barP(0,T_j) (R_{j-1,j+c}(0) \ - R_{j-1,j}(0))
\end{eqnarray}

The exponential term can be considered indeed to be a sort of
``convexity adjustment" similar in spirit to the convexity
adjustment needed to value constant maturity swaps with the LIBOR
model in the default-free market.

Finally, this formula should be tested against prices obtained via
Monte Carlo simulation of the dynamics (\ref{dRuniandbi}) before
being employed massively. One should make sure that for the order
of magnitude of volatilities, correlations and initial CDS rates
present in the market at a given time the freezing approximation
works well. We plan to analyze this approximation against Monte
Carlo simulation in further work. This future work is the reason
why we derive the exact rates dynamics below.

\end{proposition}

\begin{table}
\begin{footnotesize}
\hspace{-1cm}\begin{tabular}{|c|c|} \hline Notation & Description
\\ \hline $\tau=\tau^C$ & Default time of the reference entity ``C" \\
$T_a,(T_{a+1},\ldots,T_{b-1}), T_b$ & initial and final dates in
the protection schedule of the CDS and CMCDS\\  $T_{\beta(\tau)},
T_{\beta(t)}$ & First of the $T_i$'s following $\tau$ and $t$,
respectively \\ $\alpha_i$ & year fraction between $T_{i-1}$ and
$T_i$
\\ $L(S,T)$ & LIBOR rate at time $S$ for maturity $T$
 \\ \hline $R_{a,b}$ & Rate in the
premium leg of a CDS, paid by ``A", the protection buyer  \\
 $R_{i}$ & CDS Rate to be paid by ``A", the protection buyer at $T_i$ for protection in $[T_{i-1},T_i]$ \\
  $R_{i-2,i}$ & CDS Rate to be paid by ``A" at $T_{i-1}$ and $T_i$ for protection in $[T_{i-2},T_i]$ \\
$\rec$ & Recovery fraction on a unit notional\\
$\lgd=1-\rec$ & Protection payment against a Loss (given default of ``C" in $[T_a,T_b]$)\\
$\pprcds_{a,b}(t)$ & Discounted payoff of a postponed running CDS to ``B", the protection seller \\
$\prcds_{a,b}(t,R,\lgd)$ & Price of a running postponed CDS to ``B", protecting against default of ``C" in $[T_a,T_b]$ \\

$\cds{\mbox{\tiny CM}}_{a,b,c}(t,\lgd)$ & Price at time $t$ of a CMCDS to ``B", protecting against default  \\
    &  of ``C" in $[T_a,T_b]$ in exchange for a periodic constant maturity "$c+1$"-long CDS rate \\
\hline $1_{\{\tau > T\}}$ & Survival indicator,
is one if default occurs after $T$ and zero otherwise \\
$1_{\{\tau \le T\}}$ & Default indicator, is one if default occurs
before or at $T$, and zero otherwise
\\ $B(t)$ & Bank account numeraire of the risk neutral measure at time $t$
\\ $D(t,T)= B(t)/B(T)$ & Stochastic discount factor at time $t$ for maturity
$T$ \\
$P(t,T)$ & Zero coupon bond at time $t$ for maturity $T$\\
$1_{\{\tau > t\}}\barP(t,T)$ & Defaultable Zero coupon bond at time $t$ for maturity $T$\\
$\widehat{C}_{a,b}(t), \ \widehat{\Bbb{Q}}^{a,b}$  &  Defaultable ``Present value per basis point" numeraire and associated measure   \\
${\cal F}_t$ & Default free market information up to time $t$ \\
${\cal G}_t$ & Default free market information plus explicit
monitoring of default up to time $t$ \\
$DC(\cdot)$ & $DC(X_t)$ is the row vector $\bf{v}$ in $dX_t =
(...)dt + {\bf{v}} \ dW_t$ for diffusion processes $X$ \\  & with
$W$ \underline{vector} Brownian motion common to all relevant
diffusion processes
\\  \hline
\end{tabular}\caption{Main notation in the paper. }\label{tablenotationCDSMM}
\end{footnotesize}
\end{table}

\newpage

\section{CDS Option Market Model Dynamics}

In the remaining part of the paper we build the apparatus allowing
us to prove the main result rigorously. We specify the
probabilistic framework in the following remark.

\begin{remark}{\bf (Probabilistic Framework: $\tau$ as first jump of a Cox process)}

Here we place ourselves in a probability space $(\Omega, {\cal G},
\Qx)$ where the default time random variable $\tau$ will be
defined. The probability measure $\Qx$ is the risk neutral
probability. This space is endowed with a filtration $({\cal
F}_t)_t$, typically representing the basic filtration without
default, i.e. the ``information flow" of interest rates,
intensities and possibly other default-free market quantities.
Obviously ${\cal F}_t \subseteq {\cal G}$ for all $t$.

We consider a non-negative, $({\cal F}_t)_t$ progressively
measurable process $\lambda$ with integrable sample paths in
$(\Omega, {\cal G}, \Qx)$.

The space $(\Omega, {\cal G}, \Qx)$ is assumed to be sufficiently
rich to support a random variable $U$ uniformly distributed on
$[0,1]$ and independent of $({\cal F}_t)_t$. The random default
time $\tau$ can then be defined as \[ \tau := \inf\left\{t\ge 0:
\exp\left(-\int_{0}^t \lambda_s ds\right) \le  U \right\} \]


With this definition $\lambda$ is indeed the ${\cal F}_t$
stochastic intensity of the default time $\tau$, in that $\Qx(\tau
> t | {\cal F}_t) = \exp(-\int_{0}^t \lambda_s ds)$.

We consider also the filtration ${\cal G}_t = {\cal F}_t \vee
\sigma(\{\tau<s\}, s\le t )$ and assume ${\cal G}_t \subseteq
{\cal G}$. The second sigma field $\sigma(\{\tau<s\}, s\le t )$
contributing to ${\cal G}_t$ represents the information on whether
default occurred before $t$, and if so when exactly. Since this
information is available to us when we price, we need to condition
on ${\cal G}_t$ rather than on ${\cal F}_t$ alone.

For more details on the canonical construction of a default time
with a given hazard rate see e.g. Bielecki and Rutkowski (2001),
p. 226.
\end{remark}

The remark above amounts to saying that default $\tau$ is modeled
as the first jump time of a Cox process with the given intensity
process. We will not model the intensity directly but rather some
market quantities embedding the impact of the relevant intensity
model that is consistent with them. An explicit tractable
stochastic intensity/ interest-rate model with automatic
analytical and separable calibration to interest rate derivatives
and CDS's is given for example in Brigo and Alfonsi (2003), where
an analytical formula for CDS options based on Jamshidian's
decomposition is also presented.

Formally, we may write the RCDS discounted value at time $t$ as
\begin{eqnarray}\label{discountedpayoffcds}
\pcds_{a,b}(t) :=  \mbox{DiscountedPremiumLeg}-
\mbox{DiscountedProtectionLeg} \\ \nonumber = D(t,\tau)
(\tau-T_{\beta(\tau)-1}) R \mathbf{1}_{\{T_a < \tau < T_b \} }
\nonumber
 +   \sum_{i=a+1}^b D(t,T_i) \alpha_i R \mathbf{1}_{\{\tau \ge T_i\}
 }
 - \mathbf{1}_{\{T_a < \tau \le T_b \} }D(t,\tau) \ \lgd
\end{eqnarray}
where $t\in [T_{\beta(t)-1},T_{\beta(t)})$, i.e. $T_{\beta(t)}$ is
the first date among the $T_i$'s that follows $t$, and where
$\alpha_i$ is the year fraction between $T_{i-1}$ and $T_i$. The
stochastic discount factor at time $t$ for maturity $T$ is denoted
by $D(t,T)=B(t)/B(T)$, where $B(t)$ denotes the risk-neutral
measure bank-account numeraire.

Under the postponed formulation, where the protection payment is
moved from $\tau$ to $T_{\beta(\tau)}$, the CDS discounted payoff
can be written as
\begin{eqnarray}\label{cdspostponedpayoff}
\pprcds_{a,b}(t) := \sum_{i=a+1}^b D(t,T_i) \alpha_i R
{\mathbf{1}_{\{\tau \ge T_i\} }}
 - \sum_{i=a+1}^b \mathbf{1}_{\{T_{i-1} < \tau \le T_i \} }D(t,T_i) \
 \lgd,
\end{eqnarray}
which we term ``Postponed payments Running CDS" (PRCDS) discounted
payoff. Compare with the earlier discounted
payout~(\ref{discountedpayoffcds}) where the protection payment
occurs exactly at $\tau$: The advantage of the postponed
protection payment is that no accrued-interest term in
$(\tau-T_{\beta(\tau)-1})$ is necessary, and also that all
payments occur at the canonical grid of the $T_i$'s.

In general, we can compute the CDS price according to risk-neutral
valuation (see for example Bielecki and Rutkowski~(2001) for the
most general result of this kind):
\begin{eqnarray}\label{CDSriskneutralprice}
\cds_{a,b}(t, R, \lgd) = \Bbb{E}\left\{\pcds_{a,b}(t) |{\cal G}_t
\right\}
\end{eqnarray}
where we recall that ${\cal G}_t = {\cal F}_t  \vee
\sigma(\{\tau<u\}, u\le t )$. In the Cox process setting default
is unpredictable and this is why observation of ${\cal F}_t$ alone
does not imply observation of the default time, contrary to
standard structural (Merton, Black and Cox, etc) models where
instead ${\cal F}_t = {\cal G}_t$. At times we denote by
$\Bbb{E}_t$ and $\Bbb{Q}_t$ the expectation and probability
conditional on the default-free sigma field ${\cal F}_t$. The
above expected value can also be written as
\begin{eqnarray}\label{CDS2riskneutralprice}
\cds_{a,b}(t, R, \lgd) = \frac{\mathbf{1}_{\{\tau >
t\}}}{\Bbb{Q}(\tau>t|{\cal F}_t)} \Bbb{E}\left\{ \pcds_{a,b}(t)
|{\cal F}_t \right\}
\end{eqnarray}
(see again Bielecki and Rutkowski~(2001), or more in particular
Jeanblanc and Rutkowski (2000), where the most general form of
this result is reported).
This second expression, and especially the analogous definitions
with postponed payoffs, is fundamental for introducing the market
model for CDS options in a rigorous way. We explicit the postponed
expression by substituting the payoff: {\small
\begin{eqnarray}\label{cdspricedetails}
\prcds_{a,b}(t, R) = \frac{\mathbf{1}_{\{\tau >
t\}}}{\Bbb{Q}_t(\tau>t)} \big{\{} -\lgd \sum_{i=a+1}^b
\Bbb{E}_t[\mathbf{1}_{\{T_{i-1} < \tau \le T_i \} }D(t,T_i)] + R
\sum_{i=a+1}^b  \Bbb{E}_t [D(t,T_i) \alpha_i \mathbf{1}_{\{\tau
\ge
 T_i\}}] \big{\}}
\end{eqnarray}
}

Let us deal with the definition of postponed (running) CDS forward
rate $R_{a,b}(t)$. This can be defined as that $R$ that makes the
PRCDS value equal to zero at time $t$, so that
\[ \prcds_{a,b}(t, R_{a,b}(t),
\lgd)=0 \]

(notice that this $R_{a,b}$ is the $R^{PR}_{a,b}$ of Brigo
(2004)). The idea is then solving this equation in $R_{a,b}(t)$.
In doing this one has to be careful. It is best to start moving
from expression~(\ref{CDS2riskneutralprice}) rather
than~(\ref{CDSriskneutralprice}). Equate this expression to zero
and derive $R$ correspondingly. Strictly speaking, the resulting
$R$ would be defined on $\{\tau > t\}$ only, since elsewhere the
equation is satisfied automatically thanks to the indicator in
front of the expression, regardless of $R$. Since the value of $R$
does not matter when $\{\tau < t\}$, the equation being satisfied
automatically, we need not worry about $\{\tau < t\}$ and may
define, in general, $R=$ProtectionLegValue/PremiumLegValue,
\begin{equation}\label{Rdeffromcds} R_{a,b}(t) = \frac{\lgd\ \sum_{i=a+1}^b
\Bbb{E}[D(t,T_i)\mathbf{1}_{\{T_{i-1}< \tau\le T_i\}}|{\cal F}_t]}
{ \sum_{i=a+1}^b \alpha_i \Bbb{Q}(\tau>t|{\cal F}_t)
\bar{P}(t,T_i)}.
\end{equation}
where
\[\mathbf{1}_{\{\tau > t\}} \boxed{ \bar{P}(t,T) } :=
\Bbb{E}[D(t,T)\mathbf{1}_{\{\tau > T\}} |{\cal G}_t] =
\mathbf{1}_{\{\tau > t\}} \boxed{ \Bbb{E}[D(t,T)\mathbf{1}_{\{\tau
> T\}} |{\cal F}_t]/\Bbb{Q}(\tau>t|{\cal F}_t) }   \] is the price
at time $t$ of a defaultable zero-coupon bond maturing at time
$T$. The corresponding default free zero coupon bond is denoted by
$P(t,T)$. This approach amounts to equating to zero only the
expected value part in~(\ref{CDS2riskneutralprice}), and in a
sense is a way of privileging ``partial information" ${\cal F}_t$
expected values to ``complete information" ${\cal G}_t$ ones. Our
$R$ above is defined everywhere and not only conditional on
$\tau>t$. The technical tool allowing us to do this is the
above-mentioned Jeanblanc Rutkowski (2000) result, and this is the
spirit of part of the work in Jamshidian (2002).

A remark on how the market quotes running CDS prices is in order
at this point. First we notice that typically the $T$'s are three-
months spaced. Usually at time $t=0$, provided default has not yet
occurred, the market sets $R$ to a value $R^{\mbox{\tiny
MID}}_{a,b}(0)$ that makes the CDS fair at time $0$, i.e. such
that
$\cds_{a,b}(0,R^{\mbox{\tiny MID}}_{a,b}(0), \lgd) = 0$. Actually,
bid and ask levels are quoted for $R$. Typically in quoted CDS we
have $T_a=0$ and $T_b$ spanning a set of increasing maturities,
even though in recent times the quoting mechanism has changed in
some respects. Indeed, the quoting mechanism has become more
similar to the mechanism of the futures markets. Let $0$ be the
current time. Maturities $T_a,\ldots,T_b$ are fixed at the
original time 0 to some values such as 1y, 2y, 3y etc and then, as
time moves for example to $t= 1day$, the CDS maturities are not
shifted correspondingly of 1 day as before but remain 1y,2y etc
from the original time $0$. This means that the times to maturity
of the quoted CDS's decrease as time passes. When the quoting time
approaches maturity, a new set of maturities are fixed and so on.
A detail concerning the ``constant maturities" paradigm is that
when the first maturity $T_a$ is less than one month away from the
quoting time (say $0$), the payoff two terms
\[ T_a  D(0,T_a) R \mathbf{1}_{\{\tau > T_a \}} + (T_{a+1}-T_a) D(0,T_{a+1})  R
\mathbf{1}_{\{\tau > T_{a+1} \}}\] are replaced by
\[  T_{a+1} D(0,T_{a+1})  R
\mathbf{1}_{\{\tau > T_{a+1} \}}\] in determining the ``fair" $R$.
If we neglect this last convention, once we fix the quoting time
(say to $0$) the method to strip implied hazard functions is the
same under the two quoting paradigms. The same happens when not
neglecting the convention if we are exactly at one of the ``$0$
dates", so that for example $T_1-t=1y$. Brigo and Alfonsi~(2003)
present a more detailed section on the ``constant
time-to-maturity" earlier paradigm, and illustrate the notion of
implied deterministic intensity (hazard function).

We also set \[ \widehat{C}_{a,b}(t) := \sum_{i=a+1}^b \alpha_i
\Bbb{Q}(\tau>t|{\cal F}_t) \bar{P}(t,T_i), \] the denominator
in~(\ref{Rdeffromcds}), so that $R_{a,b}$ is a tradable asset
(upfront CDS) divided by $\widehat{C}_{a,b}(t)$. It follows that
under the measure $\widehat{\Qx}^{a,b}$ having
$\widehat{C}_{a,b}(t)$ as numeraire, $R_{a,b}$ follows a
martingale. For more details see Brigo (2004). We will be
interested in the particular cases $a=i-1,b=i$ and $a=i-2,b=i$.

In the following it will be useful to consider a running postponed
CDS on a one-period interval, with $T_a = T_{j-1}$ and $ T_b =
T_j$. We obtain, for the related CDS forward rate:
\[\hspace{0cm}  R_{j}(t) := \lgd \frac{
\Bbb{E}[D(t,T_j)\mathbf{1}_{\{T_{j-1}< \tau\le T_j\}}|{\cal F}_t]}
{ \alpha_j \Bbb{Q}(\tau>t|{\cal F}_t) \bar{P}(t,T_j)} =  \lgd
\frac{ \Bbb{E}[D(t,T_j)\mathbf{1}_{\{T_{j-1}< \tau\le T_j\}}|{\cal
F}_t]} {\widehat{C}_{j-1,j}(t) } \]
where we have set $R_j :=R_{j-1,j}$.

A last remark concerns an analogy with the default-free swap
market model, where we have a formula linking swap rates to
forward rates through a weighted average.
This is useful since it leads to an approximated formula for
swaptions in the LIBOR model, see for example Brigo and
Mercurio~(2001), Chapter~6. A similar approach  can be obtained
for CDS forward rates. It is easy to check that
\begin{equation}\label{Rasaverage}
R_{a,b}(t) = \frac{ \sum_{i=a+1}^b \alpha_i R_i(t) \barP(t,T_i)} {
\sum_{i=a+1}^b \alpha_i \bar{P}(t,T_i)}=\sum_{i=a+1}^b
\bar{w}^{a,b}_i(t) R_i(t) \approx \sum_{i=a+1}^b
\bar{w}^{a,b}_i(0) R_i(t).
\end{equation}
A possible lack of analogy with the swap rates is that the
$\bar{w}$'s
\[ \bar{w}^{a,b}_i(t) := \frac{  \alpha_i  \barP(t,T_i)} {
\sum_{h=a+1}^b \alpha_h \bar{P}(t,T_h)}  \]
 cannot be expressed as functions of the $R_i$'s only,
unless we make some particular assumptions on the correlation
between default intensities and interest rates. However, if we
freeze the $\bar{w}$'s to time $0$, which we have seen to work in
the default-free LIBOR model, we obtain easily a useful
approximated expression for $R_{a,b}$ and its volatility in terms
of $R_i$'s and their volatilities/correlations. {\bf A similar
approach is pursued in Section~\ref{oneratesonly} below, and the
reader who is interested only in the derivation of the
approximated formula given in the beginning may go there directly.
Here instead we hint at deriving the real dynamics without
compromises, which will be useful in future work for Monte Carlo
tests of the numerical approximation.}

In general, when not freezing, the presence of stochastic
intensities besides stochastic interest rates adds degrees of
freedom. Now the $\barP$'s (and thus the $\bar{w}$'s) can be
determined as functions for example of one- and two-period rates.
Indeed, it is easy to show that
\begin{equation}\label{PbarintermsofR} \barP(t,T_i) = \barP(t,T_{i-1}) \frac{\alpha_{i-1}
(R_{i-1}(t) - R_{i-2,i}(t))}{\alpha_{i} (R_{i-2,i}(t) - R_i(t))},
\ \ \frac{\bar{P}(t,T_{j})}{\bar{P}(t,T_i)} =
\frac{\alpha_{i}}{\alpha_j} \prod_{k=i+1}^{j}
\frac{R_{k-1}-R_{k-2,k}}{R_{k-2,k}-R_k   }
\end{equation}
To have the formula working we need to assume $R_{i-2,i}(t) \neq
R_i(t)$. We will therefore assume in the paper that at the initial
time $R_{i-2,i}(0) \neq R_i(0)$. Under the approximated frozen
dynamics for these two quantities we will derive below, we can see
that the probability of them to be equal at any future time $t$ is
generally $0$, since this can be computed as the probability of
the difference of two correlated continuous random variables to be
zero.

We show below how this formula helps us in obtaining a market
model for CDS rates. For the time being let us keep in mind that
the exact weights $\bar{w}(t)$ in~(\ref{Rasaverage}) are
completely specified in terms of $R_i(t)$'s and $R_{i-2,i}(t)$'s,
so that if we include these two rates in our dynamics the
``system" is closed in that we also know all the relevant
$\barP$'s. The difference with the LIBOR/Swap model is that here
to close the system we need also two-period rates.

We close this section by summarizing our notation in
Table~(\ref{tablenotationCDSMM}).

\section{One- and Two- Period CDS Rates joint Dynamics under a
single pricing measure}\label{onetworates}

Let us postulate the following dynamics for one- and two- period
CDS forward rates. Recall that $R_j = R_{j-1,j}$.

\[ d R_{j}(t) = \sigma_j(t) R_{j}(t) d Z_j^j(t) \]
\[ d R_{j-2,j}(t) = \nu_j(t;R) R_{j-2,j}(t) d V_j^{j-2,j}(t) \]

In the Brownian shocks $Z$ and $V$ the upper index denotes the
measure (i.e. the measure associated with the numeraires
$\widehat{C}_{j-1,j}, \widehat{C}_{j-2,j}$
 in the above case) and the lower index denotes to which
component of the one- and two- period rate vectors the shock
refers. {\bf The volatilities $\sigma$ are deterministic, whereas
the $\nu$'s depend on the one-period $R$'s.} We assume
correlations

\[ d Z_i d Z_j = \rho_{i,j} dt, \  \ \ d V_i d V_j = \eta_{i,j} dt,\ \  d Z_i d
V_j = \theta_{i,j} dt  \] {\bf  and $R_{i-2,i}(t) \in
(\min(R_{i-1}(t), [R_{i-1}(t)+R_i(t)]/2),\max(R_{i-1}(t),
[R_{i-1}(t)+R_i(t)]/2))$. This latter condition ensures that the
resulting $\barP$ from formula~(\ref{PbarintermsofR}) be positive
and decreasing with respect to the maturity, i.e.
$0<\barP(t,T_{i})/\barP(t,T_{i-1})<1$. The specific definition of
$\nu$ ensuring this property is currently under investigation.}

We aim at finding the drift of a generic $R_{j}$ under the measure
associated with $\widehat{C}_{i-1,i}$, let us say for $j \ge i$.

The change of numeraire toolkit provides the formula relating
shocks under $\widehat{C}_{i-1,i}$ to shocks under
$\widehat{C}_{j-2,j}$, see for example Formula (2.13) in Brigo and
Mercurio (2001), Chapter~2. We can write

\[ d \vertvect{Z^{j-2,j}}{V^{j-2,j}} = d \vertvect{Z^{i}}{V^{i}} - \mbox{CorrMatrix}
\times \mbox{VectorDiffusionCoefficient}
\left(\ln\left(\frac{\widehat{C}_{j-2,j}}{\widehat{C}_{i-1,i}}\right)\right)'
dt
\]

Let us abbreviate ``Vector Diffusion Coefficient" by ``DC".

This is actually a sort of operator for diffusion processes that
works as follows. $DC(X_t)$ is the row vector $\bf{v}$ in
\[ dX_t = (...)dt + {\bf{v}} \ d  \vertvect{Z_t}{V_t}\] for diffusion processes $X$
with $Z$ and $V$ column \underline{vectors} Brownian motions
common to all relevant diffusion processes. This is to say that if
for example $d R_1 = \sigma_1 R_1 d Z_1^1$, then
\[ DC(R_1) = [\sigma_1 R_1,\ 0,\ 0,\ldots,\ 0]. \]

Let us call $Q$ the total correlation matrix including $\rho,
\eta$ and $\theta$. We have

\[ d \vertvect{Z^{j-2,j}}{V^{j-2,j}} = d \vertvect{Z^{i}}{V^{i}} - Q
\
\mbox{DC}\left(\ln\left(\frac{\widehat{C}_{j-2,j}}{\widehat{C}_{i-1,i}}\right)\right)
dt
\]

Now we need to compute
\begin{eqnarray*}
\mbox{DC}\left(\ln\left(\frac{\widehat{C}_{j-2,j}}{\widehat{C}_{i-1,i}}\right)\right)=
\mbox{DC}\left(\ln\left(\frac{\alpha_{j-1} \bar{P}(t,T_{j-1}) +
\alpha_j \bar{P}(t,T_j)}{\alpha_i \bar{P}(t,T_i)}\right)\right)
=\\
= \mbox{DC}\left(\ln\left(\frac{\alpha_{j-1}}{\alpha_i}
\frac{\alpha_i}{\alpha_{j-1}} \prod_{k=i+1}^{j-1}
\frac{R_{k-1}-R_{k-2,k}}{R_{k-2,k}-R_k   }  +
\frac{\alpha_j}{\alpha_i} \frac{\alpha_i}{\alpha_j}
\prod_{k=i+1}^j \frac{R_{k-1}-R_{k-2,k}}{R_{k-2,k}-R_k
}\right)\right)\\= \mbox{DC}\left(\ln\left(
\left[\prod_{k=i+1}^{j-1} \frac{R_{k-1}-R_{k-2,k}}{R_{k-2,k}-R_k
}\right]\left[ 1  +
\frac{R_{j-1}-R_{j-2,j}}{R_{j-2,j}-R_j} \right]  \right)\right)\\
= \mbox{DC}\left( \sum_{k=i+1}^{j-1} \ln
\left(\frac{R_{k-1}-R_{k-2,k}}{R_{k-2,k}-R_k   }\right)\right) +
\mbox{DC}\left(\ln\left(\frac{R_{j-1}-R_{j}}{R_{j-2,j}-R_j
}\right)\right)
\\
=\sum_{k=i+1}^{j-1} \mbox{DC}\left(  \ln
\left(\frac{R_{k-1}-R_{k-2,k}}{R_{k-2,k}-R_k }\right)\right) +
\mbox{DC}\left(\ln\left(\frac{R_{j-1}-R_{j}}{R_{j-2,j}-R_j
}\right)\right) =\\ = \sum_{k=i+1}^{j-1} [\mbox{DC}  (\ln
(R_{k-1}-R_{k-2,k})) - \mbox{DC}(\ln( R_{k-2,k}-R_k ))] + \\
+ \mbox{DC}  (\ln (R_{j-1}-R_{j})) - \mbox{DC}(\ln( R_{j-2,j}-R_j
)) \\ = \sum_{k=i+1}^{j-1} \frac{\mbox{DC} (\
R_{k-1}-R_{k-2,k})}{R_{k-1}-R_{k-2,k}} - \sum_{k=i+1}^{j-1}
\frac{\mbox{DC} (\ R_{k-2,k}-R_k)}{R_{k-2,k}-R_k} + \\
+ \frac{\mbox{DC} (\ R_{j-1}-R_{j})}{R_{j-1}-R_{j}} -
\frac{\mbox{DC} (\ R_{j-2,j}-R_j)}{R_{j-2,j}-R_j} =
\\ =  \sum_{k=i+1}^{j-1}
\frac{ (\
\mbox{DC}(R_{k-1})-\mbox{DC}(R_{k-2,k}))}{R_{k-1}-R_{k-2,k}} -
\sum_{k=i+1}^{j-1} \frac{ (\
\mbox{DC}(R_{k-2,k})-\mbox{DC}(R_k))}{R_{k-2,k}-R_k}\\ + \frac{
\mbox{DC}(R_{j-1})-\mbox{DC}(R_{j})}{R_{j-1}-R_{j}} - \frac{
\mbox{DC}( R_{j-2,j})-\mbox{DC}(R_j)}{R_{j-2,j}-R_j}
\end{eqnarray*}

It follows that

\begin{eqnarray*}
d Z^{j-2,j}_m - d Z^{i}_m = - \sum_{k=i+1}^{j-1} \frac{ (
\rho_{k-1,m} \sigma_{k-1} R_{k-1} - \theta_{m,k} \nu_k
 R_{k-2,k}
)}{R_{k-1}-R_{k-2,k}} dt + \sum_{k=i+1}^{j-1} \frac{ (
\theta_{m,k} \nu_k R_{k-2,k}    - \rho_{k,m} \sigma_{k} R_{k})
 }{R_{k-2,k}- R_{k}}dt\\
- \frac{\rho_{j-1,m}\sigma_{j-1} R_{j-1} - \rho_{j,m}\sigma_{j}
R_{j} }{R_{j-1}-R_{j}}dt + \frac{\theta_{m,j} \nu_{j} R_{j-2,j} -
\rho_{j,m} \sigma_j R_j}{R_{j-2,j}-R_j}\ dt
\end{eqnarray*}

and

\begin{eqnarray*}
d V^{j-2,j}_m - d V^{i}_m = - \sum_{k=i+1}^{j-1} \frac{ (
\theta_{k-1,m} \sigma_{k-1} R_{k-1}  - \eta_{m,k} \nu_k
 R_{k-2,k}
)}{R_{k-1}-R_{k-2,k}}dt + \sum_{k=i+1}^{j-1} \frac{ ( \eta_{m,k}
\nu_k R_{k-2,k} - \theta_{k,m} \sigma_{k} R_{k})
 }{R_{k-2,k}- R_{k}} dt + \\
- \frac{\theta_{j-1,m} \sigma_{j-1} R_{j-1} - \theta_{j,m}
\sigma_{j} R_{j}  }{R_{j-1}-R_{j}}dt + \frac{\eta_{j,m}\nu_j
R_{j-2,j} - \theta_{j,m} \sigma_j R_{j}}{R_{j-2,j}-R_j}dt  =:
\bar{\phi}_m^{i,j}\ dt
\end{eqnarray*}

Therefore, by subtracting from the first equation, taking $h>i$:

\[ d Z^{h}_m  - d Z^{i}_m =   d Z^{j-2,j}_m - d Z^{i}_m - (d Z^{j-2,j}_m - d
Z^{h}_m) =\]\[= - \sum_{k=i+1}^{h} \frac{ ( \rho_{k-1,m}
\sigma_{k-1} R_{k-1} - \theta_{m,k} \nu_k
 R_{k-2,k}
)}{R_{k-1}-R_{k-2,k}} dt + \sum_{k=i+1}^{h} \frac{ ( \theta_{m,k}
\nu_k R_{k-2,k}    - \rho_{k,m} \sigma_{k} R_{k})
 }{R_{k-2,k}- R_{k}}dt = : \bar{\mu}_m^{i,h}dt  \]

so that we finally obtain (taking $h=j$)

\[ d R_{j}(t) = \sigma_j R_{j}(t)( \bar{\mu}^{i,j}_j dt + d Z_j^i(t)) \]
\[ d R_{j-2,j}(t) = \nu_j R_{j-2,j}(t))(\bar{\phi}^{i,j}_j dt + d V_j^{i}(t)), \]

or, by setting

\[ \mu^i_j := \bar{\mu}_j^{i,j}\ \sigma_j, \ \   \phi^i_j := \bar{\phi}_j^{i,j}\ \nu_j
,\]

we have

\[ d R_{j}(t) = R_{j}(t) (\mu^i_j dt  + \sigma_j  d Z_j^i(t)),
\ \  d R_{j-2,j}(t) =  R_{j-2,j}(t) (\phi^i_j dt + \nu_j d
V_j^{i}(t)),
\]

and since $\mu$ and $\phi$ are completely determined by one- and
two- period rates vectors $R = [R_{i-1,i}]_i$ and $R^{(2)} =
[R_{i-2,i}]_i$, {\bf the system is closed.} We can write a vector
SDE which is a vector diffusion for all the one- and two- period
rates under any of the $\widehat{C}_{i-1,i}$ measures:

\[ d \vertvect{R}{R^{(2)}} =  \mbox{diag}(\mu(R,R^{(2)}), \phi(R,R^{(2)})) \vertvect{R}{R^{(2)}}  dt  +
\mbox{diag}(\sigma, \nu) \vertvect{R}{R^{(2)}}\ d \vertvect{Z^i}{
V^i}
\]

At this point a Monte Carlo simulation of the process, based on a
discretization scheme for the above vector SDE is possible. One
only needs to know the initial CDS rates $R(0), R^{(2)}(0)$, which
if not directly available one can build by suitably stripping spot
CDS rates. Given the volatilities and correlations, one can easily
simulate the scheme by means of standard Gaussian shocks.

If $C$ is the Cholesky decomposition of the correlation $Q$
($Q=CC'$ with ``C" lower triangular matrix) and $W$ is a standard
Brownian motion under $\widehat{C}_{i-1,i}$, we can write

\begin{eqnarray}\label{dRuniandbi} d \vertvect{R}{R^{(2)}} =  \mbox{diag}(\mu(R,R^{(2)}),
\phi(R,R^{(2)})) \vertvect{R}{R^{(2)}}  dt  + \mbox{diag}(\sigma,
\nu) \vertvect{R}{R^{(2)}}\ C\ d W
\end{eqnarray}

The log process can be easily simulated with a Milstein scheme.

\section{An approximated model with a single family of rates and proof of the main
result}\label{oneratesonly}

Consider now the approximation
\[\hspace{0cm}  R_{j}(t) := \lgd \frac{
\Bbb{E}[D(t,T_j)\mathbf{1}_{\{T_{j-1}< \tau\le T_j\}}|{\cal F}_t]}
{ \alpha_j \Bbb{Q}(\tau>t|{\cal F}_t) \bar{P}(t,T_j)} = \lgd
\frac{ \Bbb{E}[D(t,T_{j})\mathbf{1}_{\{\tau > T_{j-1}\}}|{\cal
F}_t]-\Bbb{E}[D(t,T_j)\mathbf{1}_{\{\tau > T_{j}\}}|{\cal F}_t]} {
\alpha_j \Bbb{Q}(\tau>t|{\cal F}_t) \bar{P}(t,T_j)} = \]
\[\hspace{0cm}  = \lgd
\frac{ \Bbb{E}[D(t,T_{j-1})\mathbf{1}_{\{\tau > T_{j-1}\}}
\boxed{D(t,T_j)/D(t,T_{j-1})}|{\cal
F}_t]-\Bbb{E}[D(t,T_j)\mathbf{1}_{\{\tau
> T_{j}\}}|{\cal F}_t]} { \alpha_j \Bbb{Q}(\tau>t|{\cal F}_t)
\bar{P}(t,T_j)} =\ldots \] At this point we approximate the boxed
ratio of stochastic discount factors with the related zero-coupon
bonds, obtaining
\[ \approx \lgd
\frac{ \Bbb{E}[D(t,T_{j-1})\mathbf{1}_{\{\tau > T_{j-1}\}}|{\cal
F}_t]\boxed{P(t,T_j)/P(t,T_{j-1})}-\Bbb{E}[D(t,T_j)\mathbf{1}_{\{\tau
> T_{j}\}}|{\cal F}_t]} { \alpha_j \Bbb{Q}(\tau>t|{\cal F}_t)
\bar{P}(t,T_j)}\]\[ = \lgd \frac{\barP(t,T_{j-1})
 P(t,T_j)/P(t,T_{j-1})- \barP(t,T_j)}{ \alpha_j\bar{P}(t,T_j)}=\frac{\lgd}{\alpha_j}
 \left(\frac{\barP(t,T_{j-1})}{(1+\alpha_j F_j(t))
 \barP(t,T_j)}- 1\right)  \]\[ \approx \frac{\lgd}{\alpha_j}
 \left(\frac{\barP(t,T_{j-1})}{(1+\alpha_j F_j(0))
 \barP(t,T_j)}- 1\right) = \Rtil_j(t)\]
where $F$ is the forward LIBOR rate between $T_{j-1}$ and $T_j$.
This last definition can be inverted so as to have

\begin{equation}\label{PbarRtil}\frac{\barP(t,T_{j-1})}{\barP(t,T_{j})} = \left(
\frac{\alpha_j}{\lgd} \Rtil_j +1 \right) (1 + \alpha_j F_j(0))
> 1\end{equation}
as long as $\Rtil > 0$, provided that $F_j(0) > 0$ as should be.
This means that we are free to select any martingale dynamics for
$\Rtil_j$ under $\widehat{\Qx}^{j-1,j}$, as long as $\Rtil_j$
remains positive. Choose than such a family of $\Rtil$ as building
blocks

\[ d \Rtil_{i}(t) = \sigma_i(t) \Rtil_{i}(t) d Z_i^i(t), \ \ \mbox{for all} \ \ i \]

and define the $\barP$ by using (\ref{PbarRtil})  to obtain
inductively $\barP(t,T_j)$ from $\barP(t,T_{j-1})$ and from
$\Rtil_j$. {\bf This way, the numeraires $\barP$ become functions
only of the $\Rtil$'s, so that now the system is closed and all
one has to model is the one-period rates $\Rtil$ vector. No need
to model two-period rates in this framework.}

In this context the change of numeraire becomes

\begin{eqnarray*} d Z^{j} = d Z^{i} - \rho \mbox{DC}
\left(\ln\left(\frac{\widehat{C}_{j-1,j}}{\widehat{C}_{i-1,i}}\right)\right)'
dt = d Z^{i} - \rho  \mbox{DC}
\left(\ln\left(\frac{\barP(t,T_j)}{\barP(t,T_i)}\right)\right)' dt
=\\ =  d Z^{i} - \rho  \mbox{DC} \ln\left[\left(\prod_{h=j+1}^i
\left( \frac{\alpha_h}{\lgd} \Rtil_h +1 \right) (1 + \alpha_h
F_h(0))
  \right)\right]' dt\\ =  d Z^{i} - \rho \sum_{h=j+1}^i \mbox{DC} \ln \left(\left(
\frac{\alpha_h}{\lgd} \Rtil_h +1 \right) (1 + \alpha_h F_h(0))
  \right)' dt = \\ = d Z^{i} - \rho \sum_{h=j+1}^i \mbox{DC} \ln \left(\left(
\frac{\alpha_h}{\lgd} \Rtil_h +1 \right)
  \right)' dt = d Z^{i} - \rho \sum_{h=j+1}^i \frac{1}{\Rtil_h +\frac{\lgd}{\alpha_h}}\mbox{DC}
 (\Rtil_h)' dt
\end{eqnarray*}

so that we can write

\begin{eqnarray*} d Z^{j}_k =  d Z^{i}_k - \sum_{h=j+1}^i \rho_{k,h} \frac{\sigma_h(t) \Rtil_h}{\Rtil_h +\frac{\lgd}{\alpha_h}} dt
\end{eqnarray*}

from which we have the dynamics of $\Rtil_i$ under $\Qx^j$:

\begin{eqnarray}\label{approxfrozdynRtil} d \Rtil_{i} = \sigma_i \Rtil_i d Z_i^i = \sigma_i \Rtil_i \left(
d Z^{j}_i + \sum_{h=j+1}^i \rho_{j,h} \frac{\sigma_h
\Rtil_h}{\Rtil_h +\frac{\lgd}{\alpha_h}} dt\right) =: \Rtil_i (
\tilde{\mu}^j_i(\Rtil) dt + \sigma_i dZ_i^j)
\end{eqnarray}

Consider the drift term in the last formula. If we compute
$E^{j-1,j}[\widetilde{R}_i(T_{j-1})]$ we obtain
\begin{eqnarray}\label{meanRij-1}
E^{j-1,j}[\widetilde{R}_i(T_{j-1})] &\approx& \widetilde{R}_i(0)\exp\left\{\int^{T_{j-1}}_0\widetilde{\mu}^j_i(\widetilde{R}(0))du\right\}\nonumber\\
&=& \widetilde{R}_i(0)\exp\left\{ \sum_{k=j+1}^{i}
\frac{\widetilde{R}_k(0)}{\widetilde{R}_k(0)+\lgd /\alpha_k}
\rho_{j,k} \int_0^{T_{j-1}}\sigma_i(u) \sigma_k(u) du\right\}
\end{eqnarray}
and, if we take volatilities $\sigma$ to be constant, we have
\[\approx \widetilde{R}_i(0)\exp\left\{T_{j-1}\sigma_i\cdot\left(
\sum_{k=j+1}^{i}\rho_{j,k}\frac{\sigma_k\widetilde{R}_k(0)}{\widetilde{R}_k(0)+\lgd
/\alpha_k}\right)\right\}\]

Under independence between intensities and interest rates (and in
particular under deterministic intensities, which are a common
assumption when stripping one-period CDS rates from multi-period
ones), by definition of $R_j$ it is easy to show that at time $0$,
both the original $R_j(0)$ and the approximated $\Rtil_j(0)$ are
given in terms of the survival probabilities as
\begin{equation}
R_j(0) = \widetilde{R}_j(0)=\lgd/\alpha_j \left(\frac{\Qx(\tau
>T_{j-1})}{\Qx(\tau
>T_{j})}-1\right)
\end{equation}
and hence~\eqref{meanRij-1} is reduced to
\begin{equation}\label{meanRij-1explicit}
E^{j-1,j}[\widetilde{R}_i(T_{j-1})] = \frac{\lgd}{\alpha_i}
\left(\frac{\Qx(\tau
>T_{i-1})}{\Qx(\tau
>T_{i})}-1\right)
   \exp\left\{T_{j-1}\sigma_i
\sum_{k=j+1}^{i}\rho_{j,k}\sigma_k\left(1- \frac{\Qx(\tau
>T_{k})}{\Qx(\tau
>T_{k-1})}\right)\right\}
\end{equation}
It follows that the convexity effect vanishes if the ratio
$\frac{\Qx(\tau >T_{j})}{\Qx(\tau
>T_{j-1})}$ is close to one.

Now, based on the approximated dynamics~(\ref{approxfrozdynRtil})
and the related expectation above, we prove the main result of the
paper, i.e. Proposition~\ref{propmainrescmcds}.

\begin{proof}
To prove the proposition, we compute the price of the premium leg
as
\[ \sum_{j=a+1}^b \alpha_j \Ex_0 [ D(0,T_j) \mathbf{1}_{\{\tau > T_j \}}
R_{j-1,j+c}(T_{j-1}) ] = \ldots  \]

The first approximation we consider is~(\ref{Rasaverage}) applied
to $R_{j-1,j+c}(T_{j-1})$, so that

\[ R_{j-1,j+c}(T_{j-1}) \approx \sum_{i=j}^{j+c} \bar{w}^j_i(0)
R_i(T_{j-1}),  \ \ \bar{w}^j_i(0) = \frac{\alpha_i
\barP(0,T_i)}{\sum_{h=j}^{j+c} \alpha_h \barP(0,T_h)}\]

Then by substituting this in the premium leg expression we have
\[\ldots \approx  \sum_{j=a+1}^b \sum_{i=j}^{j+c} \alpha_j \bar{w}^j_i(0) \Ex_0 [ D(0,T_j) \mathbf{1}_{\{\tau > T_j \}}
  R_i(T_{j-1}) ] =  \]
\[ = \sum_{j=a+1}^b \sum_{i=j}^{j+c} \alpha_j \bar{w}^j_i(0) \Ex_0 [ D(0,T_j)
  R_i(T_{j-1}) \Ex( \mathbf{1}_{\{\tau > T_j \}} | {\cal F}_{T_j}) ]
  \]
\[ =\sum_{j=a+1}^b \sum_{i=j}^{j+c} \bar{w}^j_i(0) \Ex_0
\left[ \frac{B(0)}{B(T_{j})}
  (R_i(T_{j-1})\ \widehat{C}_{j-1,j}(T_j))
  \right]\]
\[ =\sum_{j=a+1}^b \sum_{i=j}^{j+c}  \bar{w}^j_i(0)\widehat{C}_{j-1,j}(0)
\widehat{\Ex}^{j-1,j}_0 [  R_i(T_{j-1}) ] = \sum_{j=a+1}^b
\sum_{i=j}^{j+c} \alpha_j \bar{w}^j_i(0) \barP(0,T_j)
\widehat{\Ex}^{j-1,j}_0 [ R_i(T_{j-1}) ] = \ldots\] where we have
applied the change of numeraire, moving from the risk neutral
numeraire $B$ to the numeraires $\widehat{C}_{j-1,j}$'s. The last
expected value can be computed based on~(\ref{meanRij-1explicit}).
By substituting the expected value expression, we obtain the final
formula.

\end{proof}

\section{A few numerical examples}

We report input data and outputs for a name with relatively large
CDS forward rates. We consider the FIAT car company CDS market
quotes as of December 20, 2004. Since in Brigo and Alfonsi (2003)
we have some evidence on the fact that CDS prices depend very
little on the correlation between interest rates and credit
spreads, when stripping credit spreads from CDS data we may assume
independence between interest rates and credit spreads. This leads
to a model where it is easy to strip default probabilities (hazard
rates) from CDS prices, as hinted at again in Brigo and Alfonsi
(2003). Using this independence assumption, we strip default (or
survival) probabilities from CDS quotes with increasing
maturities.

\subsection{Inputs}
We take as inputs the following Fiat CDS rates and use mid quotes

\vspace{1cm}

\begin{tabular}{|r|r|r|}
\hline
         $T_b$  &   $R_{0,b}^{BID} (bps) $     & $R_{0,b}^{ASK}$        \\
\hline
        1Y &       99.9 &     175.57 \\

        2Y &      172.5 &     231.38 \\

        3Y &     243.73 &     286.13 \\

        5Y &     348.85 &     366.54 \\

        7Y &        380 &        410 \\

       10Y &     395.16 &     412.73 \\
\hline
\end{tabular}

\vspace{1cm}

We take $\rec = 0.4$ (so that $\lgd = 0.6$). The input zero coupon
curve, and the survival risk-neutral probabilities stripped  from
the above CDS quotes are reported in Appendix 1 below.

\subsection{Outputs}

We start by giving a table for \[
\mbox{Conv}(\sigma,\rho):=\cmcds_{a,b,c}(0,\lgd, \sigma, \rho)-
\cmcds_{a,b,c}(0,\lgd;\rho=0).\] The first term is computed by
assuming the volatilities $\sigma_i$ of forward one-period CDS
rates $R_i$ to have a common value $\sigma$ and the pairwise
correlations $\rho_{i,j}$ to have a common value $\rho$. This
first term is then given by formula~(\ref{cmcdsvolscorr}). The
second term is the simpler value~(\ref{cmcdsnotsur}) where no
correction due to CDS forward rate dynamics is accounted for. This
difference then gives us the impact of volatilities and
correlations of CDS rates on the CMCDS price. The difference is
always positive, similarly to what happens to analogous constant
maturity swaps in default free markets under similar conditions on
volatilities and correlation. It is the impact of ``convexity" on
the CMCDS valuation. We take $a=0$, $b=20$ (5y final maturity) and
$c=21$ (which means we are considering non-standard 5y6m CDS rates
in the CMCDS premium leg, $c=19$ would amount to a 5y CDS rate).

We obtain\\

\begin{tabular}{|r|rrrr|}
\hline
$\mbox{Conv}(\sigma,\rho)$    &$\rho$: \ \         0.7 &        0.8 &        0.9 &       0.99 \\
\hline
       $\sigma$: \ \ 0.1 &   0.000659 &   0.000754 &   0.000848 &   0.000933 \\

       0.2 &   0.002662 &   0.003047 &   0.003435 &   0.003784 \\

       0.4 &   0.011066 &   0.012742 &   0.014442 &   0.015995 \\

       0.6 &   0.026619 &   0.030964 &   0.035464 &   0.039652 \\
\hline
\end{tabular}
\\

The ``convexity difference" increases with respect both to
correlation and volatility, as expected.

The next table reports the so called ``participation rate"
$\phi_{a,b,c}(\sigma,\rho)$ for a CMCDS with final $T_b=5y$ ($a=0,
b=20$, recalling that resets occur quarterly), with $5y6m$
constant maturity CDS rates ($c=21$),

\[ \phi_{0,20,21}(\sigma,\rho) = \frac{\mbox{``premium leg CDS"}}{\mbox{``premium leg CMCDS"}} = \frac{\sum_{j=1}^{20} \alpha_j \barP(0,T_j)R_{0,20}(0)}{\sum_{j=1}^{20} \alpha_j \Ex^{j-1,j}_0[ D(0,T_{j}) \mathbf{1}_{\{\tau>T_j\}}  R_{j-1,j+21}(T_{j-1})]},
 \]

The CMCDS premium leg is computed with our approximated market
model based on one-period rates~$\widetilde{R}$. As we see from
the outputs, the participation rate increases with volatility and
correlation, as is expected from the ``convexity adjustment"
effect.\\

\begin{tabular}{|r|rrrr|}
\hline
$\phi_{0,20,21}(\sigma,\rho)$    &$\rho$:\ \     0.7 &        0.8 &        0.9 &       0.99 \\
\hline
$\sigma$: \ \       0.1 &    0.71358 &    0.71325 &    0.71292 &    0.71262 \\

       0.2 &    0.70664 &    0.70532 &      0.704 &    0.70281 \\

       0.4 &    0.67894 &    0.67368 &    0.66842 &    0.66368 \\

       0.6 &    0.63302 &    0.62128 &    0.60957 &    0.59907 \\

\hline
\end{tabular}

\vspace{1cm}

Finally, we fix volatilities and correlations and check how the
patterns change when changing final maturity $T_b=T_i$. We
consider the following quantities at time $0$ and with $T_a = 0$:

\[ x_i = \frac{\mbox{``Constant maturity rate"}}{\mbox{``standard rate"}} = \frac{R_{i-1,i+c}(0)}{R_{0,b}(0)}, \ \ i=1,\ldots,b \]
\[ y_i = \frac{\Ex^{i-1,i}_0[ D(0,T_{i}) \mathbf{1}_{\{\tau>T_i\}}  R_{i-1,i+c}(T_{i-1})]}{\barP(0,T_i)R_{0,b}(0)}, \ \ i=1,\ldots,b \]
\[ z_i =\frac{\Ex^{i-1,i}_0[ D(0,T_{i}) \mathbf{1}_{\{\tau>T_i\}}
R_{i-1,i+c}(T_{i-1})]}{\barP(0,T_i) R_{i-1,i+c}(0)}, \ \ \
i=1,\ldots,b \]
\[ \psi_i = \frac{\mbox{``premium leg CDS"}}{\mbox{``premium leg CMCDS"}} = \frac{\sum_{j=1}^i \alpha_j \barP(0,T_j)R_{0,i}(0)}{\sum_{j=1}^i \alpha_j \barP(0,T_{j})  R_{j-1,j+c}(0)},
\  i=1,\ldots,b \]
\[ \phi_i = \frac{\mbox{``premium leg CDS"}}{\mbox{``premium leg CMCDS with convexity"}}=\frac{\sum_{j=1}^i \alpha_j \barP(0,T_j)R_{0,i}(0)}{\sum_{j=1}^i \alpha_j \Ex^{j-1,j}_0[ D(0,T_{j}) \mathbf{1}_{\{\tau>T_j\}}
R_{j-1,j+c}(T_{j-1})]}.
 \]

\vspace{1cm}

\begin{tabular}{|rrrrr|r|}
\hline        $x_i$ &        $y_i$ &        $z_i$ &      $\psi_i$ &      $\phi_i$ &  \\
\hline              &              &              &               &               &  $\sigma$ = 0.4; \\
           &            &            &            &            & $\rho$ = 0.9; \\

    1.0668 &     1.0668 &          1 &    0.37773 &    0.37773 &            \\

    1.1288 &     1.1359 &     1.0063 &    0.36281 &    0.36162 & $\rec$ = 0.4; \\

    1.1914 &     1.2075 &     1.0135 &    0.35281 &    0.35039 &       a=0; \\

    1.2525 &     1.2792 &     1.0214 &    0.34359 &    0.33993 &    c = 20; \\

    1.3107 &     1.3495 &     1.0297 &    0.33512 &    0.33024 &    b = 20; \\

    1.3673 &     1.4193 &      1.038 &    0.34187 &    0.33548 &            \\

    1.4171 &     1.4826 &     1.0462 &    0.36905 &    0.36064 &            \\

    1.4515 &       1.53 &     1.0541 &    0.40755 &    0.39664 &            \\

    1.4716 &     1.5622 &     1.0616 &    0.45262 &    0.43881 &            \\

    1.4798 &     1.5818 &     1.0689 &    0.49477 &    0.47785 &            \\

    1.4837 &     1.5979 &     1.0769 &    0.52661 &    0.50671 &            \\

    1.4905 &     1.6175 &     1.0852 &    0.55072 &    0.52799 &            \\

    1.4999 &     1.6403 &     1.0936 &    0.56931 &    0.54384 &            \\

    1.5122 &      1.666 &     1.1018 &    0.58674 &    0.55846 &            \\

    1.5236 &       1.69 &     1.1092 &    0.60704 &    0.57574 &            \\

    1.5275 &      1.706 &     1.1168 &    0.62715 &     0.5928 &            \\

    1.5274 &     1.7174 &     1.1244 &    0.64681 &    0.60938 &            \\

    1.5249 &     1.7236 &     1.1303 &    0.67017 &    0.62939 &            \\

    1.5106 &     1.7173 &     1.1368 &    0.69254 &    0.64843 &            \\

    1.4924 &     1.7047 &     1.1422 &    0.71589 &    0.66842 &
    \\
\hline
\end{tabular}

\vspace{1cm}

The $x_i$'s measure how the constant maturity CDS rate differs
multiplicatively from the standard CDS rate, so they are a measure
of how the constant maturity CDS differs from a standard CDS in
the premium rate paid at each period. We find an increasing
pattern in $T_i$ as partly expected from the fact that the input
CDS rates $R^{BID,ASK}_{0,b}$ are increasing with respect to
maturity $T_b$.

The $y_i$'s measure the same effect while taking into account
``convexity", i.e. future randomness of the payoff and
correlation. The $y_i$'s would reduce to the $x_i$'s if
correlations $\rho$ were taken equal to $0$. The $y$ maintain the
increasing pattern with respect to $T_i$.

The $z_i$'s measure the multiplicative impact of ``convexity", in
that they are due to contributions stemming from volatilities
$\sigma$ and correlations $\rho$ of CDS rates. The impact is
increasing with maturity $T_i$, as expected from the sign in the
exponent of the convexity adjustments and from the positive signs
of correlations (and volatilities).

Finally, as seen above, the $\psi_i$'s are the so called
``participation rates" for different terminal maturities $T_i$.
They give the ratio between the premium leg in  a standard CDS
protecting in $[0,T_i]$ and the premium leg in  CMCDS for the same
protection interval when ignoring the convexity adjustment due to
correlation and volatilities. The $\phi_i$'s are the participation
rates computed when taking into account convexity due to
volatilities and correlations. We have seen a particular
participation rate $\phi$ earlier. In the table above for $\phi_i$
we obtain an initially decreasing pattern followed by a longer
increasing pattern for both $\psi$ and $\phi$. Notice that, on the
longest participation rate, in the last row of the related table,
convexity has an impact moving from a $71.59\%$ participation rate
when not including ``convexity" (ignoring correlations and
volatilities) to a $66.84\%$ participation rate when including
convexity. There is a $4.8\%$ difference in the participation rate
of this FIAT 5y-5y3m CMCDS with correlations set at 0.9 and
volatilities at $40\%$.

\section{Further work}
In further research we need to propose a realistic dynamics for
one- and two- period rates that completely specifies the market
model, along the guidelines given in this paper and in Brigo
(2004). Then we may test the market formula proposed here against
Monte Carlo simulation of the exact dynamics. Moreover, examining
the formula outputs for $\rho$ matrices with more realistic
decorrelation patterns and for different names can be
appropriated.


\section*{Appendix 1. Input data and survival probabilities}

The input zero coupon curve, and the survival risk-neutral
probabilities stripped  from FIAT CDS quotes are:

\vspace{1cm}


\begin{tabular}{|r|r|r|r|}
\hline
    $\alpha_i$ &         $T_i$ &    $P(0,T_i)$ &    $\Qx(\tau > T_i)$ \\
\hline
         0 &          0 &    0.99994 &    0.99994 \\

   0.24444 &    0.24444 &    0.99459 &    0.99429 \\

   0.25556 &        0.5 &      0.989 &    0.98856 \\

   0.25556 &    0.75556 &    0.98309 &    0.98279 \\

   0.25278 &     1.0083 &    0.97709 &    0.97712 \\

      0.25 &     1.2583 &    0.97098 &    0.97155 \\

   0.25556 &     1.5139 &    0.96458 &    0.96433 \\

   0.25556 &     1.7694 &      0.958 &    0.95409 \\

   0.25278 &     2.0222 &    0.95133 &    0.94108 \\

      0.25 &     2.2722 &    0.94448 &    0.92552 \\

   0.25556 &     2.5278 &     0.9373 &     0.9086 \\

   0.25556 &     2.7833 &    0.92995 &    0.89227 \\

   0.25278 &     3.0361 &    0.92251 &    0.87669 \\

   0.25278 &     3.2889 &    0.91502 &    0.86165 \\

   0.25556 &     3.5444 &    0.90731 &    0.84618 \\

   0.25556 &        3.8 &    0.89929 &    0.82931 \\

   0.25278 &     4.0528 &    0.89139 &    0.81203 \\

      0.25 &     4.3028 &    0.88373 &    0.79449 \\

   0.25556 &     4.5583 &    0.87544 &    0.77495 \\

   0.25556 &     4.8139 &     0.8673 &    0.75531 \\

   0.25278 &     5.0667 &    0.85906 &    0.73503 \\

      0.25 &     5.3167 &    0.85085 &     0.7142 \\

   0.25556 &     5.5722 &    0.84255 &    0.69403 \\

   0.25556 &     5.8278 &    0.83417 &    0.67559 \\

\hline
\end{tabular}

\vspace{1cm}

(continues in the next page)

\newpage

\begin{tabular}{|r|r|r|r|}
\hline
    $\alpha_i$ &         $T_i$ &    $P(0,T_i)$ &    $\Qx(\tau > T_i)$ \\

   0.25278 &     6.0806 &    0.82572 &    0.65879 \\

      0.25 &     6.3306 &    0.81735 &    0.64351 \\

   0.25556 &     6.5861 &    0.80892 &    0.62968 \\

   0.25556 &     6.8417 &    0.80035 &    0.61709 \\

   0.25278 &     7.0944 &    0.79182 &    0.60594 \\

   0.25278 &     7.3472 &    0.78344 &    0.59601 \\

   0.25556 &     7.6028 &    0.77494 &    0.58651 \\

   0.25556 &     7.8583 &    0.76641 &    0.57685 \\

   0.25278 &     8.1111 &    0.75794 &    0.56715 \\

      0.25 &     8.3611 &    0.74977 &    0.55744 \\

   0.25556 &     8.6167 &    0.74141 &     0.5474 \\

   0.25556 &     8.8722 &    0.73303 &    0.53726 \\

   0.25278 &      9.125 &    0.72474 &    0.52713 \\

      0.25 &      9.375 &     0.7168 &    0.51705 \\

   0.25556 &     9.6306 &    0.70869 &    0.50667 \\

   0.25556 &     9.8861 &    0.70041 &    0.49601 \\

   0.25278 &     10.139 &    0.69241 &    0.48565 \\

      0.25 &     10.389 &     0.6849 &     0.4756 \\
\hline
\end{tabular}

\end{document}